\documentclass[aps,prd,reprint,twocolumn,superscriptaddress,longbibliography,nofootinbib,floatfix,showpacs]{revtex4-2}

\usepackage{graphicx}

\usepackage{amsmath,amssymb,amsfonts}
\usepackage{hyperref}
\usepackage{slashed}
\usepackage{bm}
\usepackage[usenames]{color}

\newcommand{\be}{\begin{equation}}
\newcommand{\ee}{\end{equation}}
\newcommand{\ud}{\mathrm{d}}

\newcommand{\LCm}{{\scriptscriptstyle -}} 
\newcommand{\LCp}{{\scriptscriptstyle +}}
\newcommand{\LCperp}{{\scriptscriptstyle \perp}}

\usepackage[T1]{fontenc} \usepackage[latin1]{inputenc}

\begin{document}

\title{Momentum correlation in pair production by spacetime dependent fields from scattered wave functions}
\author{Greger Torgrimsson}
\email{greger.torgrimsson@umu.se}
\affiliation{Department of Physics, Ume{\aa} University, SE-901 87 Ume{\aa}, Sweden}

\begin{abstract}

We consider Sauter-Schwinger pair production by electric fields that depend on both time and space, $E(t,z)$ and $E(t,x,y)$. For space-independent fields, $E(t)$, momentum conservation, $\delta({\bf p}+{\bf p}')$, fixes the positron momentum, ${\bf p}'$, in terms of the electron momentum, ${\bf p}$. For $E(t,z)$, on the other hand, $p_z$ and $p'_z$ are independent. However, previous exact-numerical studies have considered only the probability as a function of a single momentum variable, $P(p_z)$, $P(p'_z)$ or $P(p'_z-p_z)$, but not the correlation $P(p_z,p'_z)$. In this paper, we show how to obtain $P(p_z,p'_z)$ by solving the Dirac equation numerically. To do so, we split the wave function into a background and a scattered wave, $\psi(t,{\bf x})=\psi_{\rm back.}(t,{\bf x})+\psi_{\rm scat.}(t,{\bf x})$, where $\psi_{\rm back.}\propto\exp(\pm ipx+\text{gauge term})$. $\psi_{\rm scat.}$ vanishes outside a past light cone and is obtained by solving $(i\slashed{D}-m)\psi_{\rm scat.}=-(i\slashed{D}-m)\psi_{\rm back.}$ backwards in time starting with $\psi_{\rm scat.}(t\to+\infty,{\bf x})=0$.         

\end{abstract}

\maketitle

\section{Introduction}

The vast majority of papers on Schwinger pair production~\cite{Dunne:2004nc,Fedotov:2022ely} focus on 1D fields, $E(t)$, $E(z)$ or $E(t+z)$. However, in the past decade or so, there have been a considerable number of papers on 2D fields~\cite{Hebenstreit:2011wk,Jiang:2012mfn,Jiang:2013wha,Wollert:2015kra,Kohlfurst:2015zxi,Kohlfurst:2015niu,Aleksandrov:2016lxd,Aleksandrov:2017mtq,Kohlfurst:2017hbd,Kohlfurst:2017git,Lv:2018wpn,Peng:2018hmj,Ababekri:2019dkl,Kohlfurst:2019mag,Aleksandrov:2019ddt,Ababekri:2019qiw,Li:2021vjf,Mohamedsedik:2021pzb,Li:2021wag,Kohlfurst:2021skr,Jiang:2023hbo}, e.g. $E(t,z)$. Since there is no momentum conservation in the $z$ direction, the probability, $P(p,p')$, depends independently on the electron and positron momenta, $p_z$ and $p_z'$. However, previous studies have only considered quantities where one momentum variable has been integrated out, e.g. the probability, $P(p)=\int\ud p'P(p,p')$, or the average number of electrons with a given momentum, $N(p)$. One reason for this reduction is practical; previously used methods give $P(p)$ or $P(p-p')$ directly, while $P(p,p')$ appears nowhere in the calculations. Reducing the number of parameters also makes it easier to scan the parameter space. 

However, there are patterns in the correlation $P(p,p')$ that cannot be seen in $P(p)$. In~\cite{DegliEsposti:2022yqw,DegliEsposti:2023qqu,DegliEsposti:2024upq} we showed how to use worldline instantons~\cite{Affleck:1981bma,Dunne:2005sx,Dunne:2006st,Dunne:2006ur,Schneider:2018huk} to obtain a weak-field approximation of $P(p,p')$. In~\cite{DegliEsposti:2024upq} we found that fields with more than two peaks can lead to 2D moir\'e patterns in $P(p,p')$. Such patterns disappear if one integrates out $p$ or $p'$. In this paper, we will show how to obtain the exact $P(p,p')$ (though still to zeroth order in the Furry-picture expansion) by solving the Dirac equation numerically.

\section{Derivation}

The first steps in the following derivation are standard and can be found e.g. in~\cite{unstableVacuumBook}. 
We are interested in the Dirac equation in a background field $A_\mu(x^\nu)$,
\be\label{DiracPsi}
(i\slashed{D}-1)\psi=0 \;,
\ee
where $D_\mu=\partial_\mu+iA_\mu$, $c=\hbar=m=1$ and we have rescaled $eA_\mu\to A_\mu$, so to zeroth order in the Furry picture there are no explicit factors of $e$. 
The inner product of two solutions of the Dirac equation is denoted
\be\label{innerProduct}
(\psi_1,\psi_2)=\int\ud^3{\bf x}\psi_1^\dagger\psi_2 \;.
\ee
The solution of~\eqref{DiracPsi} is expanded in terms of basis states for an incoming or outgoing electron or positron, $U_{\rm in}$, $V_{\rm in}$, $U_{\rm out}$ and $V_{\rm out}$, where
\be\label{UVasympLineA}
\begin{split}
\lim_{t\to-\infty}U_{\rm in}(s{\bf p}x)&=U_{-\infty}(s{\bf p}x) \\
\lim_{t\to-\infty}V_{\rm in}(s{\bf p}x)&=V_{-\infty}(s{\bf p}x) \\
\lim_{t\to\infty}U_{\rm out}(s{\bf p}x)&=U_\infty(s{\bf p}x) \\
\lim_{t\to\infty}V_{\rm out}(s{\bf p}x)&=V_\infty(s{\bf p}x) \;,
\end{split}
\ee
where, in the temporal gauge $A_0=0$,
\be\label{UVback}
\begin{split}
U_{\tilde{t}}(s{\bf p}x^\mu)&=u_s({\bf p})\exp\left[-ipx-i\int^{\bf x}\!\!\ud y^k A_k(\tilde{t},{\bf y})\right] \\
V_{\tilde{t}}(s{\bf p}x^\mu)&=v_s({\bf p})\exp\left[ipx-i\int^{\bf x}\!\!\ud y^k A_k(\tilde{t},{\bf y})\right]
\end{split}
\ee
and the spinors are normalized such that
\be
\begin{split}
(U_{\rm in}[s,{\bf p}],U_{\rm in}[r,{\bf q}])&=(2\pi)^3\delta^3({\bf p}-{\bf q})\delta_{sr}\\
(V_{\rm in}[s,{\bf p}],V_{\rm in}[r,{\bf q}])&=(2\pi)^3\delta^3({\bf p}-{\bf q})\delta_{sr} \\
(U_{\rm in}[s,{\bf p}],V_{\rm in}[r,{\bf q}])&=0 \;,
\end{split}
\ee
and similarly for the out solutions. The Dirac operator can be expressed in either the in or out basis,
\be\label{modeExp}
\begin{split}
\Psi&=\int\frac{\ud^3{\bf q}}{(2\pi)^3}\sum_r\left[U_{\rm in}(r{\bf q}x)a_{\rm in}(r{\bf q})+V_{\rm in}(r{\bf q}x)b_{\rm in}^\dagger(r{\bf q})\right]\\
&=\int\frac{\ud^3{\bf q}}{(2\pi)^3}\sum_r\left[U_{\rm out}a_{\rm out}+V_{\rm out}b_{\rm out}^\dagger\right] \;,
\end{split}
\ee
where the mode operators obey
\be\label{acommutator}
\{a_{\rm in}(r{\bf q}),a_{\rm in}^\dagger(s{\bf p})\}=(2\pi)^3\delta({\bf q}-{\bf p})\delta_{rs}
\qquad\text{etc.} \;,
\ee
the completeness of these states means
\be\label{completeness}
\int\frac{\ud^3{\bf q}}{(2\pi)^3}\sum_r\Big[U_{\rm in}(x)U_{\rm in}^\dagger(y)+V_{\rm in}(x)V_{\rm in}^\dagger(y)\Big]=\delta({\bf x}-{\bf y}) \;,
\ee
and
\be
\{\Psi(t,{\bf x}),\bar{\Psi}(t,{\bf y})\}=\gamma^0\delta({\bf x}-{\bf y}) \;.
\ee

We obtain the Bogoliubov coefficients by projecting~\eqref{modeExp} with the basis functions,
\be
\begin{split}
a_{\rm in}(s{\bf p})=\int\frac{\ud^3{\bf q}}{(2\pi)^3}\sum_r\Big[&(U_{\rm in}[s{\bf p}],U_{\rm out}[r{\bf q}])a_{\rm out}(r{\bf q})\\
+&(U_{\rm in}[s{\bf p}],V_{\rm out}[r{\bf q}])b_{\rm out}^\dagger(r{\bf q})\Big] \;.
\end{split}
\ee
By viewing $s$ and ${\bf p}$ as the components of an infinite-dimensional vector $a_{\rm in}(s{\bf p})$, and $(U_{\rm in}[s{\bf p}],U_{\rm out}[r{\bf q}])$ as a matrix, we can write the Bogoliubov transformation in a more compact way as in~\cite{unstableVacuumBook}, with an implicit sum
\be\label{inFromOut}
\begin{split}
a_{\rm in}&=(U_{\rm in}|U_{\rm out})a_{\rm out}+(U_{\rm in}|V_{\rm out})b^\dagger_{\rm out}\\
b^\dagger_{\rm in}&=(V_{\rm in}|U_{\rm out})a_{\rm out}+(V_{\rm in}|V_{\rm out})b^\dagger_{\rm out} \;,
\end{split}
\ee
and
\be\label{outFromIn}
\begin{split}
a_{\rm out}&=(U_{\rm out}|U_{\rm in})a_{\rm in}+(U_{\rm out}|V_{\rm in})b^\dagger_{\rm in}\\
b^\dagger_{\rm out}&=(V_{\rm out}|U_{\rm in})a_{\rm in}+(V_{\rm out}|V_{\rm in})b^\dagger_{\rm in} \;.
\end{split}
\ee
The completeness relation~\eqref{completeness} can be expressed as
\be\label{uuvvone}
|U_{\rm in})(U_{\rm in}|+|V_{\rm in})(V_{\rm in}|={\bf 1} \;.
\ee

From~\eqref{inFromOut} we have
\be\label{normalOrder}
\begin{split}
a_{\rm out}&=(U_{\rm in}|U_{\rm out})^{-1}[a_{\rm in}-(U_{\rm in}|V_{\rm out})b^\dagger_{\rm out}]\\
b_{\rm out}&=[b_{\rm in}-a^\dagger_{\rm out}(U_{\rm out}|V_{\rm in})](V_{\rm out}|V_{\rm in})^{-1} \;,
\end{split}
\ee
where $(U_{\rm in}|U_{\rm out})^{-1}$ is the inverse of $(U_{\rm in}|U_{\rm out})$.
With~\eqref{normalOrder} one can calculate the probability amplitude for pair production~\cite{unstableVacuumBook},
\be\label{pairAmp}
\begin{split}
M_{mn}&={}_{\rm out}\langle0|a_{\rm out}(m)b_{\rm out}(n)|0\rangle_{\rm in}\\
&={}_{\rm out}\langle0|0\rangle_{\rm in}[(U_{\rm in}|U_{\rm out})^{-1}(U_{\rm in}|V_{\rm out})]_{mn}\\
&=-{}_{\rm out}\langle0|0\rangle_{\rm in}[(U_{\rm out}|V_{\rm in})(V_{\rm out}|V_{\rm in})^{-1}]_{mn}\;,
\end{split}
\ee
where $m$ and $n$ are indices for both spin and momentum. 

From~\eqref{outFromIn}, the expectation value of electrons in state $m$ is given by~\cite{unstableVacuumBook}
\be\label{Nem}
\begin{split}
N_{e^\LCm}(m)&={}_{\rm in}\langle0|a_{\rm out}^\dagger(m)a_{\rm out}(m)|0\rangle_{\rm in}\\
&=[(U_{\rm out}|V_{\rm in})(V_{\rm in}|U_{\rm out})]_{mm} \;,
\end{split}
\ee
and similarly for the expectation value of positrons
\be\label{Nep}
\begin{split}
N_{e^\LCp}(m)&={}_{\rm in}\langle0|b_{\rm out}^\dagger(m)b_{\rm out}(m)|0\rangle_{\rm in}\\
&=[(V_{\rm out}|U_{\rm in})(U_{\rm in}|V_{\rm out})]_{mm} \;.
\end{split}
\ee

With the mode operators normalized as in~\eqref{acommutator}, the total number of pairs is given by
\be\label{NemSum}
N_{e^\LCm}=\sum_m N_{e^\LCm}(m)=\int\frac{\ud^3{\bf p}}{(2\pi)^3}\sum_s N_{e^\LCm}(s{\bf p}) \;.
\ee

The pair-production amplitude~\eqref{pairAmp} gives us full access to the correlation between the electron and positron, but is complicated because of the inverse $(...|...)^{-1}$. The expectation values do not involve any inverse, but do not give any information about the correlation. Since previous studies have focused on $N_{e^\LCm}(m)$ or $N_{e^\LCp}(m)$, this explains why, as far as we know, the correlation has not previously been studied using an exact numerical treatment. 

Another popular method for studying pair production is the Dirac-Heisenberg-Wigner method~\cite{Bialynicki-Birula:1991jwl}, where the central object is the Wigner function, $W(t,{\bf x},\Delta{\bf p})$, where $\Delta{\bf p}={\bf p}'-{\bf p}$. At each moment in time, $W(t,{\bf x},\Delta{\bf p})$ is a function of $2n$ variables, where $n$ is the number of nontrivial spatial dimensions. This is the same number of variables as $P({\bf p},{\bf p}')$. However, although $W(t,{\bf x},\Delta{\bf p})$ has been computed for $F_{\mu\nu}(t,z)$ in several papers, as far as we know, no one has tried to extract the momentum correlation from $W(t,{\bf x},\Delta{\bf p})$. Instead, the numerical result for $W(t,{\bf x},\Delta{\bf p})$ has been integrated over ${\bf x}$, leaving a function of a single momentum variable, $\Delta{\bf p}$. We will not consider the Wigner method in this paper, because of memory concerns: At each time step, the computer would need to handle $W(t,{\bf x},\Delta{\bf p})$ on a grid in both ${\bf x}$ and $\Delta{\bf p}$, which consequently becomes very large for multidimensional fields. In contrast, when solving the Dirac equation directly, one can consider each value of ${\bf p}$ and ${\bf p}'$ separately, so one only needs a grid in ${\bf x}$.

\subsection{Momentum correlation}

The correlation can be studied without any inverse by considering  
\be
\begin{split}
N_{e^\LCm e^\LCp}(m,n)&={}_{\rm in}\langle0|a_{\rm out}^\dagger(m)a_{\rm out}(m)b_{\rm out}^\dagger(n)b_{\rm out}(n)|0\rangle_{\rm in}\\
&=N_{e^\LCm}(m)N_{e^\LCp}(n)+N_1(m,n) \;,
\end{split}
\ee
where $N_1$ can be written in different ways with the help of~\eqref{uuvvone},
\be\label{N1}
\begin{split}
N_1&=-[(U_{\rm out}|U_{\rm in})(U_{\rm in}|V_{\rm out})]_{mn}[(V_{\rm out}|V_{\rm in})(V_{\rm in}|U_{\rm out})]_{nm}\\
&=[(U_{\rm out}|U_{\rm in})(U_{\rm in}|V_{\rm out})]_{mn}[(V_{\rm out}|U_{\rm in})(U_{\rm in}|U_{\rm out})]_{nm}\\
&=|[(U_{\rm out}|U_{\rm in})(U_{\rm in}|V_{\rm out})]_{mn}|^2\\
&=|[(U_{\rm out}|V_{\rm in})(V_{\rm in}|V_{\rm out})]_{mn}|^2\;.
\end{split}
\ee
Computing both the last two lines and comparing them is one way to check the precision of the numerical results.

We also have $N_1=|\phi|^2$, where
\be
\phi={}_{\rm in}\langle0|a_{\rm out}(m)b_{\rm out}(n)|0\rangle_{\rm in} \;.
\ee
The correlator $\phi$ appears in some papers on Schwinger pair production by space independent fields~\cite{Schmidt:1998vi,Kim:2011jw}.

For $E<1$ the probability is small, and this suppression comes from factors of $(U|V)$. There are two factors of $(U|V)$ in $N_1$ but four factors in $N_{e^\LCm}(m)N_{e^\LCp}(n)$. Therefore, in a regime where one needs to include contributions which are only suppressed by powers of $E$, but where exponentially suppressed terms are still negligible, one has $N_{e^\LCm e^\LCp}\approx N_1$ to a good approximation. In any case, $N_{e^\LCm}(m)$ and $N_{e^\LCp}(m)$ can be obtained as byproducts in the computation of $N_1$. 

In this regime, we can also obtain~\eqref{N1} from~\eqref{pairAmp} by noting that ${}_{\rm out}\langle0|0\rangle_{\rm in}\approx1$ and, using~\eqref{uuvvone},
\be
\begin{split}
{}_m(V_{\rm out}|V_{\rm in})(V_{\rm in}|V_{\rm out})_n&=\delta_{mn}-{}_m(V_{\rm out}|U_{\rm in})(U_{\rm in}|V_{\rm out})_n\\
&\approx\delta_{mn} \;,
\end{split}
\ee
so $(V_{\rm out}|V_{\rm in})^{-1}\approx(V_{\rm in}|V_{\rm out})$ and hence $|\eqref{pairAmp}|^2\approx\eqref{N1}$.

\begin{figure*}
    \centering
    \includegraphics[width=0.49\linewidth]{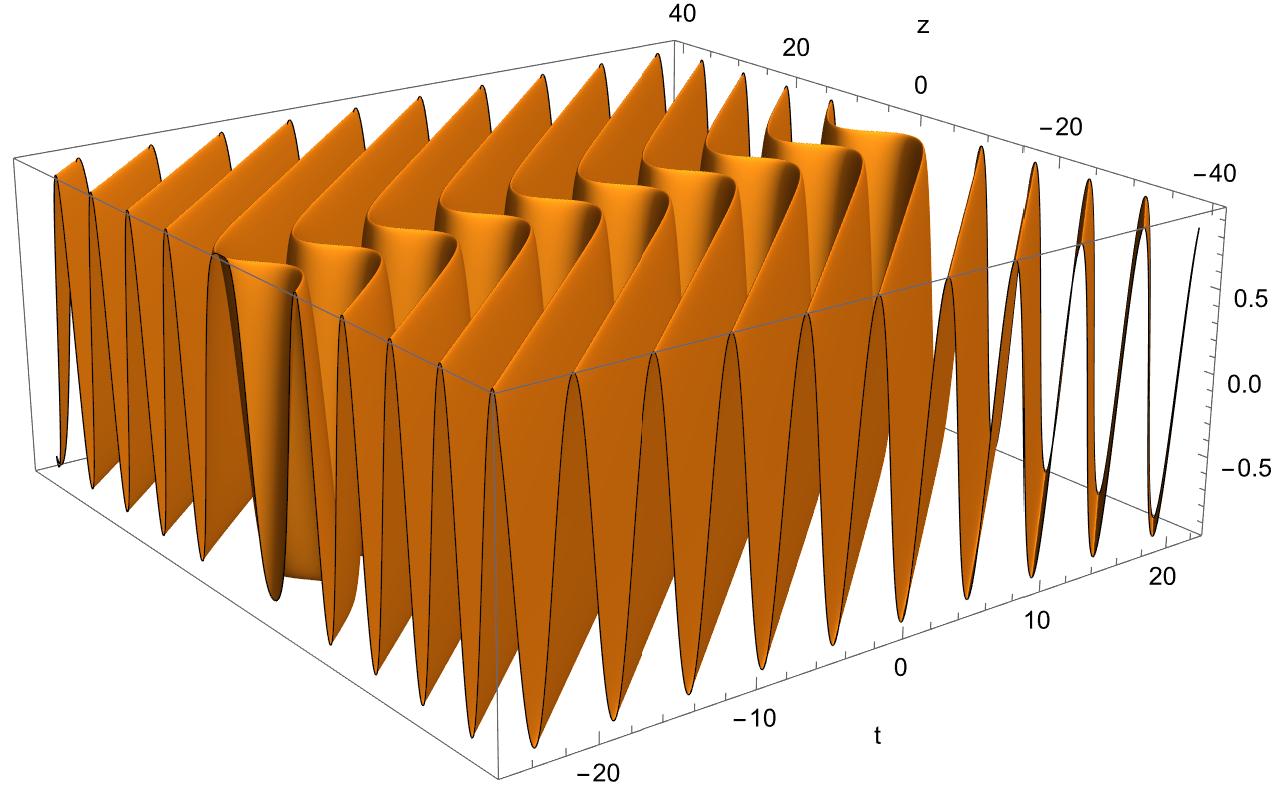}
    \includegraphics[width=0.49\linewidth]{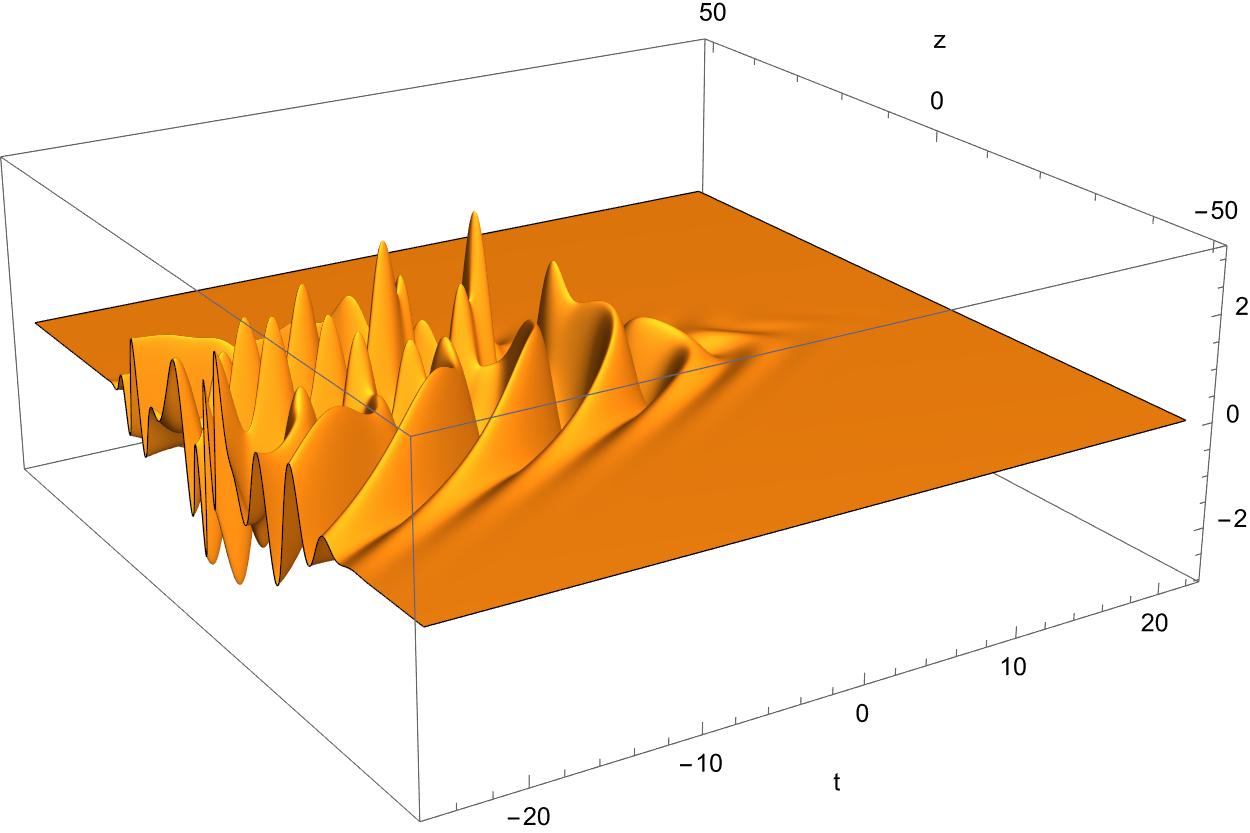}
    \caption{Background wave function (left), $U_\infty$, and scattered wave function (right), $\Delta U$, in~\eqref{UVsplit}. The plots show the real part of the first component of the 2D spinors in Sec.~\ref{2D fields}. The field is $A_3(t,z)=(E/\omega)\tanh(\omega t)\text{sech}^2(\kappa z)$ with $E=1/3$ and $\omega=\kappa=2E/3$. For this gauge choice, $A_3(\infty,z)\ne0$, so $U_\infty$ is not just a plane wave, but includes the gauge integral in the exponent. $\Delta U$ is obtained by solving~\eqref{DiracDelta2D} backwards in time, starting with $\Delta U=0$ in the asymptotic future, $t=t_{\rm out}\sim22$. The momentum, $p_3=-0.85$, enters~\eqref{DiracDelta2D} via  $U_\infty$. The pair probability is obtained by Fourier transforming the $z$ dependence of $\psi(t,z)$ in the asymptotic past, $t=t_{\rm in}\sim-25$.}
    \label{fig:split}
\end{figure*}

\subsection{Splitting the wavefunction}

We consider background fields that are negligible outside a finite space-time region. To compute $\psi(t,{\bf x})$ we start at $t=t_{\rm out}$, where $t_{\rm out}$ is large enough so that $F_{\mu\nu}(t>t_{\rm out},{\bf x})\approx0$. We use~\eqref{UVasympLineA} as ``initial'' conditions for $U_{\rm out}(s,{\bf p},x^\mu)$ and $V_{\rm out}(s',{\bf p}',x^\mu)$ at $t_{\rm out}$, integrate backwards in time to some $t_{\rm in}$, chosen so that $F_{\mu\nu}(t<t_{\rm in},{\bf x})\approx0$, and then project $U_{\rm out}(t_{\rm in})$ and $V_{\rm out}(t_{\rm in})$ onto $U_{\rm in}(r,{\bf q},t_{\rm in},{\bf x})$ or $V_{\rm in}(r,{\bf q},t_{\rm in},{\bf x})$ and sum over $r$ and ${\bf q}$. 

We solve the Dirac equation by splitting the wave function into a background wave function and a scattered wave function, 
\be
\psi=\psi_{\rm back.}+\psi_{\rm scat.} \;,
\ee
as illustrated in Fig.~\ref{fig:split}. Since $F_{\mu\nu}(t,{\bf x})$ has (approximately) finite support and since $\psi_{\rm scat.}$ cannot propagate faster than the speed of light, $\psi_{\rm scat.}(t>t_{\rm in},z)$ is only nonzero inside the past light cone of $F_{\mu\nu}(t,{\bf x})$. Outside this light cone, we have $\psi\approx\psi_{\rm back}$, and in particular $U_{\rm out}\approx U_\infty$ and $V_{\rm out}\approx V_\infty$. We therefore choose $\psi_{\rm back}=U_\infty$ for $\psi=U_{\rm out}$ and $\psi_{\rm back}=V_\infty$ for $\psi=V_{\rm out}$. We denote the corresponding scattered waves as $\Delta U$ and $\Delta V$. Thus,
\be\label{UVsplit}
\begin{split}
U_{\rm out}(s,{\bf p},x^\mu)&=U_\infty(s,{\bf p},x^\mu)+\Delta U(s,{\bf p},x^\mu) \\   
V_{\rm out}(s,{\bf p},x^\mu)&=V_\infty(s,{\bf p},x^\mu)+\Delta V(s,{\bf p},x^\mu) \;,
\end{split}
\ee
where $U_\infty$ and $V_\infty$ are given by~\eqref{UVback}, while $\Delta U$ and $\Delta V$ are obtained by solving an inhomogenous equation with $U_\infty$ and $V_\infty$ as source terms,
\be\label{DiracDelta}
\begin{split}
(i\slashed{D}-1)\Delta U&=-(i\slashed{D}-1)U_\infty=[\slashed{A}(t,{\bf x})-\slashed{A}(\infty,{\bf x})]U_\infty \\
(i\slashed{D}-1)\Delta V&=-(i\slashed{D}-1)V_\infty=[\slashed{A}(t,{\bf x})-\slashed{A}(\infty,{\bf x})]V_\infty
\end{split}
\ee
with ``initial'' conditions after the background field has vanished,
\be
\Delta U(t_{\rm out},{\bf x})=0 \qquad \Delta V(t_{\rm out},{\bf x})=0 \;.
\ee
Inside the lightcone, the line integral in~\eqref{UVback} may depend on the path, but whatever path we choose, there is a corresponding dependence in $\Delta U$ and $\Delta V$, so that the observables are path independent.

We evaluate all the inner products in~\eqref{N1} at $t=t_{\rm in}$, where $U_\infty$ and $V_\infty$ are given by the in solutions, 
\be\label{U0V0t0}
\begin{split}
U_\infty(s{\bf p}x^\mu)&=e^{i\mathcal{P}({\bf x})}U_{-\infty}(s{\bf p}x^\mu) \\
V_\infty(s{\bf p}x^\mu)&=e^{i\mathcal{P}({\bf x})}V_{-\infty}(s{\bf p}x^\mu) \;,
\end{split}
\ee
except for an Aharonov-Bohm phase,
\be
\mathcal{P}({\bf x})=\int^{\bf x}\!\!\ud y^k[A_k(-\infty,{\bf y})-A_k(\infty,{\bf y})] \;.
\ee

\subsection{Vanishing Aharonov-Bohm phase, $\mathcal{P}=0$}

${\bf A}(-\infty,{\bf x})={\bf A}(\infty,{\bf x})$ (or simply ${\bf A}(\pm\infty,{\bf x})=0$) is a particularly relevant case from a phenomenological point of view. Then $\mathcal{P}=0$ and we can simplify~\eqref{N1} using
\be
\begin{split}
{}_m(U_{\rm out}|U_{\rm in})_l&={}_m(U_\infty|U_{\rm in})_l+{}_m(\Delta U|U_{\rm in})_l \\
&=\delta_{ml}+{}_m(\Delta U|U_\infty)_l \\
{}_l(U_{\rm in}|V_{\rm out})_n&={}_l(U_\infty|\Delta V)_n \;,
\end{split}
\ee
where all the inner products are evaluated at the same\footnote{Note that $(\psi_1|\psi_2)$ is time independent if $\psi_1$ and $\psi_2$ are solutions to the Dirac equation, but $\Delta U$ and $\Delta V$ alone are not solutions.} $t=t_{\rm in}$. We find
\be\label{N1compact}
\begin{split}
N_1&=\big|{}_m(U_\infty|\Delta V)_n+{}_m(\Delta U|U_\infty)(U_\infty|\Delta V)_n\big|_{t=t_{\rm in}}^2 \\
&=\big|{}_m(\Delta U|V_\infty)_n+{}_m(\Delta U|V_\infty)(V_\infty|\Delta V)_n\big|_{t=t_{\rm in}}^2 
\end{split}
\ee
and
\be
\begin{split}
N_{e^\LCm}(m)&={}_m(\Delta U|V_\infty)(V_\infty|\Delta U)_m \\
N_{e^\LCp}(m)&={}_m(\Delta V|U_\infty)(U_\infty|\Delta V)_m \;.
\end{split}
\ee
Thus, we have formulated everything in terms of ${\bf x}$ integrals over $\Delta U$ and $\Delta V$, which have (approximately) finite support.

\subsection{$\mathcal{P}(z)\ne0$}

Below, we will consider examples of $A_3(t,z)$ with $\mathcal{P}({\bf x})=\mathcal{P}(z)\ne0$. For $(U_{\rm in}[q]|U_\infty[p])$ we find the following integral,
\be\label{Jdef}
J(\Delta)=\int_{-\infty}^\infty\ud z\exp\left[i\Delta z+i\mathcal{P}(z)\right] \;,
\ee
where $\Delta=q_3-p_3$. For $\mathcal{P}=0$ we have $J=2\pi\delta(\Delta)$. For $\mathcal{P}\ne0$ we can separate $\mathcal{J}$ into terms with and without $\delta(\Delta)$ as follows.
We split the integral into three pieces,
\be
\int_{-\infty}^{z_\LCm}+\int_{z_\LCm}^{z_\LCp}+\int_{z_\LCp}^\infty \;,
\ee
where $z_\LCm$ and $z_\LCp$ are chosen such that $\mathcal{P}(z<z_\LCm)\approx\mathcal{P}(-\infty)$ and $\mathcal{P}(z>z_\LCp)\approx\mathcal{P}(\infty)$. In the first integral we assume an integration contour for $q_3$ equivalent to $\Delta\to\Delta-i\epsilon$, with $0<\epsilon\ll1$, and for the last integral $\Delta\to\Delta+i\epsilon$. By using
\be
\frac{1}{\Delta\pm i\epsilon}=\mp i\pi\delta(\Delta)+\text{p.v.}\frac{1}{\Delta}
\ee
and performing partial integration on the $z_\LCm<z<z_\LCp$ integral, we find
\be\label{Jdeltapv}
\begin{split}
J(\Delta)&=2\pi\delta(\Delta)\frac{1}{2}\left[e^{i\mathcal{P}(-\infty)}+e^{i\mathcal{P}(\infty)}\right]\\
&-\text{p.v.}\frac{1}{\Delta}\int_{-\infty}^\infty\ud z\,\mathcal{P}'(z)\exp\left\{i\Delta z+i\mathcal{P}(z)\right\}\\
&=:2\pi\delta(\Delta)\mathcal{E}-\text{p.v.}\frac{\hat{J}(\Delta)}{\Delta}\;.
\end{split}
\ee
Now all the $z$ integrals have finite support.  

Another way to derive~\eqref{Jdeltapv} is to write~\eqref{Jdef} as $J(\Delta)=L'(\Delta)$, where
\be
\begin{split}
L(\Delta)&=\int_0^\infty\ud z\bigg\{i\left[e^{i\mathcal{P}(-z)}-e^{i\mathcal{P}(z)}\right]\frac{\cos(\Delta z)}{z}\\
&+\left[e^{i\mathcal{P}(-z)}+e^{i\mathcal{P}(z)}\right]\frac{\sin(\Delta z)}{z}\bigg\} \;,
\end{split}
\ee
and perform partial integration using $\cos/z=\text{Ci}'(z)$ and $\sin/z=\text{Si}'(z)$. The boundary term, $\pi\text{sign}(\Delta)\mathcal{E}$, gives the $\delta(\Delta)$ term after differentiating $L(\Delta)$. 

The factor $\text{p.v.}1/\Delta$ is harmless from an analytical point of view, but needs to be treated carefully to avoid losing precision in the numerics. This is an issue that one does not encounter for $N_{e^\LCm}$ and $N_{e^\LCp}$, since it comes from $(U|U)$ or $(V|V)$ in the correlation~\eqref{N1}. There are different ways of dealing with this, the most straightforward being 
\be
\text{p.v.}\int_{-\infty}^\infty\frac{\ud\Delta}{\Delta}f(\Delta)=\int_0^\infty\frac{\ud\Delta}{\Delta}[f(\Delta)-f(-\Delta)] \;.
\ee
Some precision may be lost in $f(\Delta)-f(-\Delta)$, which could be a problem since $f$ is proportional to $(V|U)$ which is typically relatively small at $\Delta=0$, i.e. $q_3=p_3$, so, were it not for the $1/\Delta$ factor, one would have been content with a lower precision for $f$ at $\Delta=0$.

\subsection{Spinors}\label{spinor section}

The spinors in~\eqref{UVasympLineA} can be chosen as
\be\label{uv}
u_s({\bf p})=\frac{(1+\slashed{p})R_s}{\sqrt{2p_0(p_0+\epsilon p_3)}}
\quad
v_s({\bf p})=\frac{(1-\slashed{p})R_s}{\sqrt{2p_0(p_0+\epsilon p_3)}} \;,
\ee
where 
\be\label{eigenR}
\gamma^0\gamma^3R_s=\epsilon R_s=\pm R_s 
\qquad
i\gamma^1\gamma^2R_s=sR_s=\pm R_s
\ee
and normalization constants chosen so that
\be
u^\dagger_r({\bf p})u_s({\bf p})=v^\dagger_r({\bf p})v_s({\bf p})=R^\dagger_r R_s=\delta_{rs} \;.
\ee
$\epsilon=1$ gives one basis for both spin states for both electrons and positrons, and $\epsilon=-1$ gives another basis. So, $\epsilon$ is not a spin index to be summed over. We could do all calculations with either $\epsilon=1$ or $\epsilon=-1$. 

In the chiral basis,
\be
\gamma^0=\begin{pmatrix}0&1\\1&0\end{pmatrix}
\qquad
\gamma^j=\begin{pmatrix}0&\sigma_j\\-\sigma_j&0\end{pmatrix} \;,
\ee
where $\sigma_j$ are the Pauli matrices,
\be
\sigma_1=\begin{pmatrix}0&1\\1&0\end{pmatrix}
\quad
\sigma_2=\begin{pmatrix}0&-i\\i&0\end{pmatrix}
\quad
\sigma_3=\begin{pmatrix}1&0\\0&-1\end{pmatrix} \;,
\ee
we have
\be
\begin{split}
\epsilon=1:& \qquad R_\LCm=(0,1,0,0) \qquad R_\LCp=(0,0,1,0)\\
\epsilon=-1:& \qquad R_\LCp=(1,0,0,0) \qquad R_\LCm=(0,0,0,1) \;.
\end{split}
\ee

Without using any particular representation, we find
\be\label{RRzero}
\begin{split}
R_r^\dagger\gamma^0 R_s&=\epsilon R_r^\dagger\gamma^0\gamma^0\gamma^3R_s
=-\epsilon R_r^\dagger\gamma^0\gamma^3\gamma^0R_s\\
&=-\epsilon^2R_r^\dagger\gamma^0 R_s=0   \;, 
\end{split}
\ee
and similarly $R_r^\dagger\gamma^3 R_s=R_r^\dagger\gamma^0\gamma^{1,2}R_s=R_r^\dagger\gamma^3\gamma^{1,2}R_s=0$.

\subsection{Symmetric fields}

For symmetric fields, ${\bf A}(t,-{\bf x})={\bf A}(t,{\bf x})$, there is a symmetry between electrons and positrons. If $\psi(t,{\bf x})$ is a solution to~\eqref{DiracPsi}, then so too is
\be
\Tilde{\psi}(t,{\bf x})=C\psi^*(t,-{\bf x}) \;,
\ee
where $\psi^*$ is the complex conjugate (not the Hermitian/conjugate transpose) and
\be
C(\gamma^0)^*C^{-1}=-\gamma^0 \qquad C(\gamma^k)^*C^{-1}=\gamma^k \;.
\ee
$C$ is the usual charge-conjugation matrix~\cite{Itzykson:1980rh}, but the charge conjugation of $\psi$ would be $\psi^c(t,{\bf x})=C\bar{\psi}^T(t,{\bf x})$. In the chiral basis we could choose $C=\gamma^0\gamma^2$.
From the exponents in~\eqref{UVasympLineA} we see that
\be\label{chargeConjugation}
\begin{split}
V_{\rm in}(t,{\bf x},{\bf p})&=CU_{\rm in}^*(t,-{\bf x},-{\bf p}) \\
V_{\rm out}(t,{\bf x},{\bf p})&=CU_{\rm out}^*(t,-{\bf x},-{\bf p}) \;,
\end{split}
\ee
hence
\be
(V_{\rm in}[{\bf q}]|V_{\rm out}[{\bf p}'])=(U_{\rm in}[-{\bf q}]|U_{\rm out}[-{\bf p}'])^* 
\ee
and
\be
\begin{split}
&\int\ud^3{\bf q}(U_{\rm out}[{\bf p}]|V_{\rm in}[{\bf q}])(V_{\rm in}[{\bf q}]|V_{\rm out}[{\bf p}'])\\
=&\int\ud^3{\bf q}\big\{(U_{\rm in}[{\bf q}]|U_{\rm out}[-{\bf p}'])(V_{\rm in}[-{\bf q}]|U_{\rm out}[{\bf p}])\big\}^* \;.
\end{split}
\ee
Thus, for symmetric fields and ${\bf p}'=-{\bf p}$, we only need to compute $U_{\rm out}$, while in general we need to independently compute both $U_{\rm out}$ and $V_{\rm out}$.

\subsection{2D fields}\label{2D fields}

In this section we will consider $A_3(t,z)$ and $p_\LCperp=0$. $\gamma^1$ and $\gamma^2$ drop out, and it follows from~\eqref{RRzero} that
\be
u_r^\dagger f(\gamma^0,\gamma^3)u_s\propto\delta_{rs} \qquad\text{etc.}
\ee
so one can reduce the 4D spinors to 2D, where the following gamma matrices act as Pauli matrices,
\be
-\gamma^0\gamma^3\to\sigma_1
\quad i\gamma^3\to\sigma_2
\quad \gamma^0\to\sigma_3 \;.
\ee
The Dirac equation reduces to~\cite{Jiang:2023hbo}
\be\label{Dirac2D}
i\partial_t\psi=(iD_3\sigma_1+\sigma_3)\psi \;.
\ee
The electron and positron ``spinors'' are given by
\be
u(q_3)=\frac{(1+q_0,q_3)}{\sqrt{2q_0(q_0+1)}}
\ee
\be
v(q_3)=\sqrt{\frac{1+q_0}{2q_0}}\left(\frac{q_0-1}{q_3},1\right) \;.
\ee
In this 2D space, $C$ in~\eqref{chargeConjugation} reduces to an irrelevant phase.

Splitting the wave functions as in~\eqref{UVsplit} gives
\be\label{DiracDelta2D}
\begin{split}
i\partial_t\psi_{\rm scat.}&=([i\partial_3-A_3(t,z)]\sigma_1+\sigma_3)\psi_{\rm scat.}\\
&+[A_3(\infty,z)-A_3(t,z)]\sigma_1\psi_{\rm back.} \;.
\end{split}
\ee

Since we consider fields which do not depend on $x$ and $y$, we have
\be
(V[{\bf q}]|U[{\bf p}])_{4D}=(2\pi)^2\delta^2(q_\LCperp+p_\LCperp)(V[{\bf q}]|U[{\bf p}])_{2D} \;.
\ee
Summing over spins and $q_\LCperp$ in~\eqref{NemSum} gives
\be\label{NemSumtz}
\sum_sN_{e^\LCm}^{4D}(s{\bf p})=2V_\LCperp N_{e^\LCm}^{2D}({\bf p}) \;,
\ee
where $V_\LCperp=V_1V_2=(2\pi)^2\delta_\LCperp^2(0)$ and $N_{e^\LCm}^{2D}$ is $N$ without $\int\ud^2q_\LCperp/(2\pi)^2$, $(2\pi)^2\delta_\LCperp^2(...)$, $\delta_{rs}$ and $\sum_s$. $N_1$ has an overall $(2\pi)^2\delta_\LCperp^2(p+p')$, so we integrate over the transverse momentum of the positron, $p_\LCperp'$,
\be
\begin{split}
&\int\frac{\ud^2 p_\LCperp'}{(2\pi)^2}\left|\int\frac{\ud^2 q}{(2\pi)^2}(2\pi)^2\delta_\LCperp^2(p-q)(2\pi)^2\delta_\LCperp^2(q+p')\right|^2\\
&=V_\LCperp \;.
\end{split}
\ee
Thus,
\be\label{N12D}
N_1({\bf p}p'_3)=\sum_{ss'}\int\frac{\ud^2 p_\LCperp'}{(2\pi)^6} N_1(s{\bf p}s'{\bf p}')
=\frac{2V_\LCperp}{(2\pi)^4}N_1^{2D} \;,
\ee
where $N_1^{2D}$ is $N_1(s{\bf p}s'{\bf p}')$ without all the trivial stuff.

\subsection{Numerical approach}

The Dirac equation can be conveniently solved using pseudospectral methods~\cite{Jiang:2012mfn,Jiang:2013wha,Lv:2018wpn,Jiang:2023hbo,Kohlfurst:2015zxi,Kohlfurst:2017hbd,Kohlfurst:2017git,Kohlfurst:2019mag,Antoine:2019fwz}. We solve~\eqref{DiracDelta2D} by discretizing $z$ into $m\gg1$ points,
\be
z=z_0+(z_1-z_0)\frac{j-1}{m-1} \qquad j=1,2,\dots m \;,
\ee
where $-z_0$ and $z_1$ should be large enough so that both $E(t,z)$ and $\psi_{\rm scat.}(t>t_{\rm in},z)$ are contained in the interval $(z_0,z_1)$. We choose for simplicity $z_0=-z_1$ and $m$ an odd integer. The spatial derivatives in~\eqref{DiracDelta2D} are computed by fast-Fourier-transforming (FFT), $\psi_{\rm scat.}(t,z)\to\psi_{\rm scat.}(t,k)$, multiplying by the wave vector $k$, and then FFT back from $k$ to $z$ space. 
Since $\psi_{\rm scat.}(t>t_{\rm in},z)$ is negligible outside $(z_0,z_1)$, one can pad the list $\psi_{\rm scat.}(t,z_j)$ with zeros to obtain a denser grid in $k$,
\be
k=\frac{2\pi}{n\Delta z}j
\quad \Delta z=\frac{z_1-z_0}{m-1}
\quad -\frac{n-1}{2}\leq j\leq \frac{n-1}{2} \;,
\ee
where $n-m\geq0$ is the number of padded zeros.

In Mathematica, this can be done as follows. At each time step, the spatial derivative is computed as
\be
\begin{split}
\varphi p&=\text{ArrayPad}\left[\varphi,\frac{n-m}{2}\right] \\
d\varphi&=-i\text{ InverseFourier}[k\text{ Fourier}[\varphi p]]\\
d\varphi&=\text{ArrayPad}\left[d\varphi,-\frac{n-m}{2}\right]\;,
\end{split}
\ee
where $k$ is a list of the $k$ values (appropriately sorted) and $\varphi$ is an $m$-dimensional list of $\psi_1(t,z_j)$ or $\psi_2(t,z_j)$. The time evolution is solved as a set of $2m$ coupled ordinary differential equation using
\be
\begin{split}
\psi s=\text{NDSolveValue}[&\{\psi'[t]\resizebox{2mm}{\height}{==}F[t,\psi[t]],\psi[\text{tOut}]\resizebox{2mm}{\height}{==}\psi\text{Out}\}\\
&,\psi,\{t,\text{tIn},\text{tOut}\}] \;,
\end{split}
\ee
where $F$ takes a number ($t$) and an $(m,m)$-dimensional array ($\psi[t]$) and outputs the right-hand-side of $-i\eqref{DiracDelta2D}$, and
\be
\psi\text{Out}=\{\text{Table}[0,m],\text{Table}[0,m]\} \;.
\ee
From the discretized solution, $\psi s[\text{tIn}]$, we can obtain a smooth function using Interpolation[...]. Projecting the solution onto $U_{\rm in}$ and $V_{\rm in}$ can be done with Fourier[...] and NIntegrate[...].

\begin{figure}
    \centering
    \includegraphics[width=\linewidth]{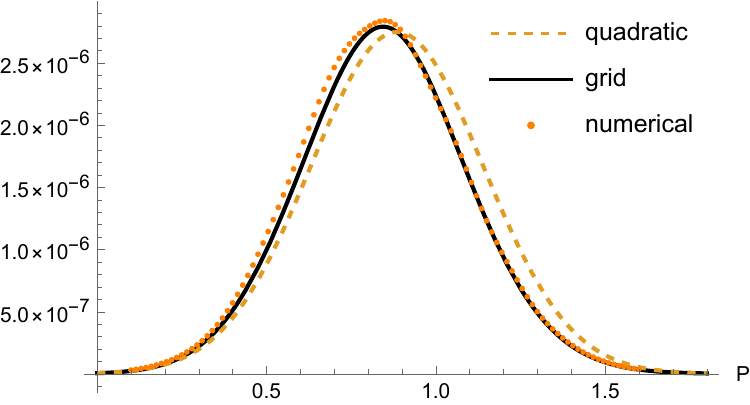}
    \includegraphics[width=\linewidth]{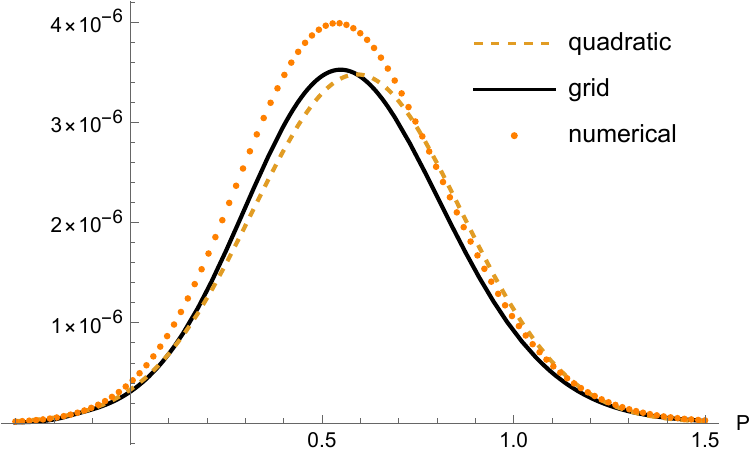}
    \caption{The spectrum for $E(t,z)=E_0\text{sech}^2(\omega t)\text{sech}^2(\kappa z)$ with $E_0=1/3$ and $\gamma_\omega=\gamma_\kappa$, where $\gamma_\omega=\omega/E_0$ and $\gamma_\kappa=\kappa/E_0$. $\gamma=2/3$ in the first plot, and $\gamma=1$ in the second. The plots show a cross section of the spectrum, $\eqref{N12D}/V_\LCperp$, where $p_3=-P+\Delta/2$ and $p'_3=P+\Delta/2$ with $\Delta=0$. The ``numerical'' result has been obtained by solving~\eqref{DiracDelta2D} numerically. Each data point takes about 3 seconds to compute on a laptop, with a grid in $z$ with $\mathcal{O}(200)$ points. The ``grid'' and ``quadratic'' lines show the worldline-instanton ($E\ll1$) approximations, obtained with the methods described in~\cite{DegliEsposti:2024upq}.}
    \label{fig:singlePeak}
\end{figure}

\begin{figure}
    \centering
    \includegraphics[width=\linewidth]{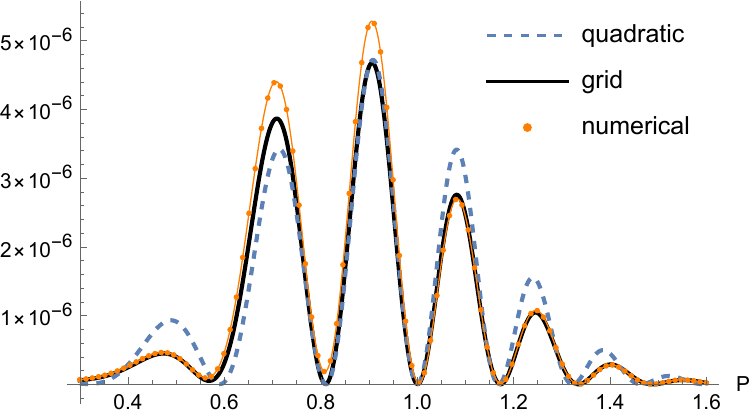}
    \caption{Spectrum for $E(t,z)=E_0[\text{sech}^2(\omega t)+\text{sech}^2(\omega t-5)]\text{sech}^2(\kappa z)$ with $E_0=0.3$ and $\gamma_\omega=\gamma_\kappa=2/3$. Same notation as in Fig.~\ref{fig:singlePeak}. Each numerical point took about one minute to compute, with $\mathcal{O}(700)$ grid points. From the instanton approximations for similar examples shown in Fig.~11 in~\cite{DegliEsposti:2024upq}, one expects essentially no oscillations in the $\Delta$ direction for this type of superposition of two pulses separated in $t$ but not in $z$.}
    \label{fig:twoPeaksP}
\end{figure}

\begin{figure}
    \centering
    \includegraphics[width=\linewidth]{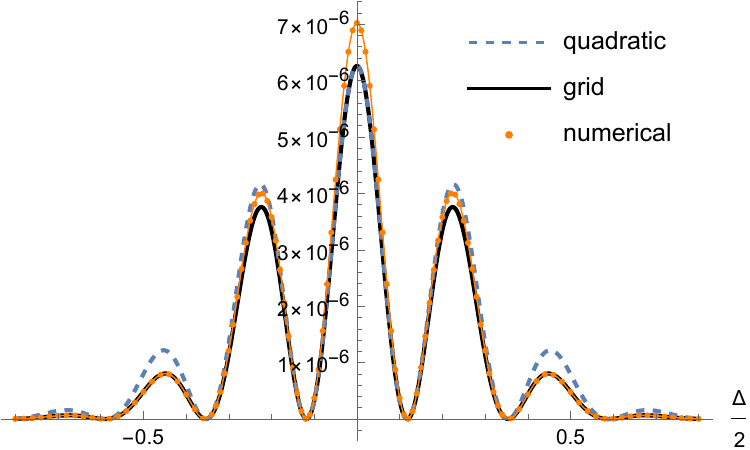}
    \caption{Spectrum for $E(t,z)=E_0\text{sech}^2(\omega t)[\text{sech}^2(\kappa z+2)+\text{sech}^2(\kappa z-2)]$ with $E_0=\omega=\kappa=0.3$. The plot shows a cross section of the spectrum, $p_3=-P+\Delta/2$ and $p'_3=P+\Delta/2$ with $P=P_{\rm saddle}\approx0.59$. From the instanton approximations for similar examples shown in Fig.~6 in~\cite{DegliEsposti:2024upq}, one expects essentially no oscillations in the $P$ direction for this type of superposition of two pulses separated in $z$ but not in $t$.}
    \label{fig:twoPeaksDelta}
\end{figure}

\begin{figure}
    \centering
    \includegraphics[width=\linewidth]{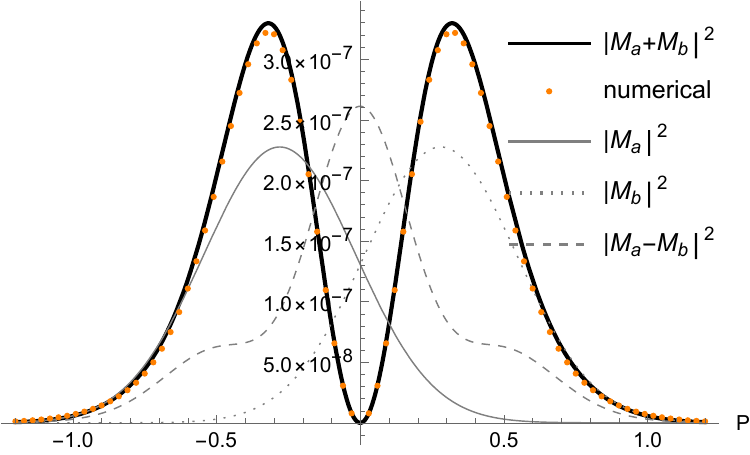}
    \caption{Spectrum for $E(t,z)=2E_0\kappa z\exp[-(\kappa z)^2-(\omega t)^2]$ with $E_0=\omega=\kappa=1/3$. This field has two peaks with opposite signs, $E_{\rm min}=E(0,z\approx-0.7/\kappa)<0$ and $E_{\rm max}=E(0,z\approx0.7/\kappa)>0$. The plot shows a cross section with $\Delta=0$. $|M_a+M_b|^2$ gives what we in the other plots refer to as the quadratic approximation. For this particular example, $M_a$ is the amplitude obtained by expanding around an instanton with a turning point ($\dot{t}(0)=0$) near $z(0)\approx-0.75$ and with asymptotic momentum $P_a\approx-0.28$ and $\Delta_a\approx0.11$. $M_b$ corresponds to $z(0)\approx0.75$, $P_b\approx0.28$ and $\Delta_b\approx-0.11$. $|M_a-M_b|^2$ shows what one would have obtained if one had made a mistake in calculating the relative sign of the two amplitude terms. The agreement between $|M_a+M_b|^2$ and the numerical results confirms that we have the correct sign.}
    \label{fig:zGauss}
\end{figure}

\section{Numerical examples}

To check these methods, we compare with the results obtained using the instanton approximation~\cite{DegliEsposti:2024upq}. Consider the superposition of two pulses,
\be\label{twoPulses}
E_3(t,z)=E^{(1)}(t-t_a,z-z_a)+E^{(1)}(t-t_b,z-z_b) \;,
\ee
where $t_{a,b}$ and $z_{a,b}$ are some constants, and $E^{(1)}(t,z)$ some single-peak pulse, e.g. a Sauter pulse,
\be\label{SauterPulse}
E^{(1)}(t,z)=E_0\text{sech}^2(\omega t)\text{sech}^2(\kappa z) \;.
\ee
Figs.~\ref{fig:singlePeak}, \ref{fig:twoPeaksP}, \ref{fig:twoPeaksDelta} and~\ref{fig:zGauss} show a couple of examples. The results are in good agreement with the instanton approximations from~\cite{DegliEsposti:2024upq}. In fact, in several cases the agreement is much better than what one might have expected, given that the instanton results give the leading order in a $E\ll1$ expansion and $E=0.3$ or $E=1/3$ is not particularly small. One should therefore expect larger errors in general. 

\begin{figure}
    \centering
    \includegraphics[width=\linewidth]{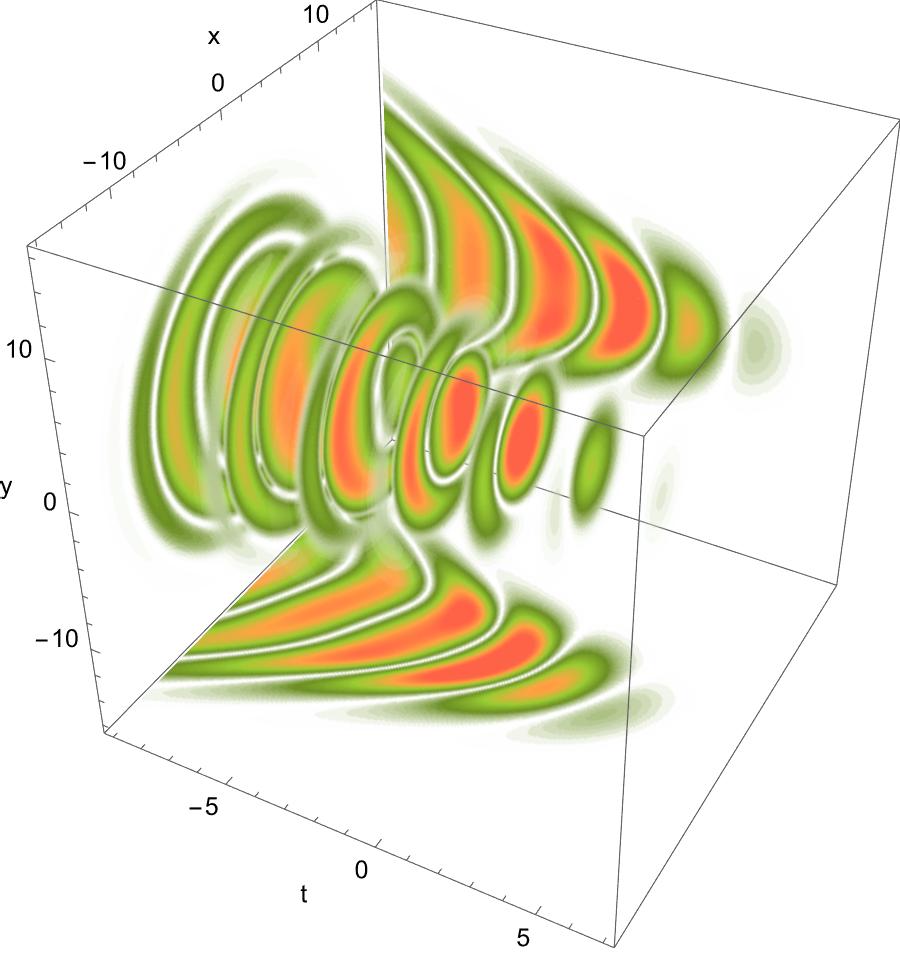}
    \caption{The scattered electron wave function, $\Delta U$, for the field in~\eqref{A0txy} with $E_0=\omega=\kappa=1/3$ and with momentum $(p_1,p_2)=(-0.35,0)$. The density plots on the $t=(-9,-6.75,-4.5,-2.25,0,2.25,4.5)$ planes show $f=|\text{Re }\Delta U_1|$, where the color runs from white/transparent to green to orange/red as the value of $f$ goes from $0$ to the maximum value of $f$. The wall at constant $x$ shows $f=|\text{Re }\Delta U_1(t,x=0,y)|$. The floor shows $f=|\text{Re }\Delta U_1(t,x,y=0)|$.}
    \label{fig:threeD}
\end{figure}

\section{$2+1$-dimensional fields}

In this section we will consider fields which are nontrivial in 3 dimensions, $A_\mu(t,x,y)$ and $A_3=0$, so the electric field lies in the plane, ${\bf E}=E_x{\bf e}_x+E_y{\bf e}_y$, while the magnetic field is orthogonal, ${\bf B}=B{\bf e}_z$. We also assume that $p_3=p'_3=0$. As is well known, in $2+1$-dimensions one can use $2\times2$ dimensional gamma matrices, e.g.
\be
\gamma^0=\sigma_3
\quad
\gamma^1=i\sigma_2
\quad
\gamma^2=-i\sigma_1 \;.
\ee
The Dirac equation reduces to
\be
\partial_0\psi=-(\sigma_1D_1+\sigma_2D_2+i\sigma_3+iA_0)\psi \;,
\ee
and the free spinors are given by
\be
\begin{split}
u&=\frac{1}{\sqrt{2p_0(p_0+1)}}(1+p_0,-p_1-ip_2)\\
v&=\frac{1}{\sqrt{2p_0}}\left(\frac{\sqrt{p_0-1}}{\sqrt{p_1^2+p_2^2}}(-p_1+ip_2),\frac{\sqrt{p_1^2+p_2^2}}{\sqrt{p_0-1}}\right) \;.
\end{split}
\ee
We split the wave function $\psi=\psi_{\rm back.}+\psi_{\rm scat.}$ as before. Now we also assume fields for which $A_\mu\to0$ in all asymptotic directions, so that we can use~\eqref{N1compact}. 

As a simple example, we consider
\be\label{A0txy}
A_0(t,x,y)=\frac{E_0}{\kappa}\exp\left[-(\omega t)^2-(\kappa x)^2-(\kappa y)^2\right] \;.
\ee
The lack of momentum conservation means that $p_1$, $p_2$, $p'_1$ and $p_2'$ are all independent parameters. For a cylindrically symmetric field, such as~\eqref{A0txy}, the momentum spectrum also has a symmetry which reduces the size/dimension of the part of momentum space that needs to be plotted, which reduces the number of times we need to solve the Dirac equation. However, for any given values of $p_1$, $p_2$, $p'_1$ and $p_2'$, the method we have used to solve the Dirac equation does not assume such a symmetry. 

Fig.~\ref{fig:threeD} shows one example of $\Delta U$. It was obtained using a grid for $-22<x,y<22$ with $201\times201$ points, and $401\times401$ points in the Fourier space. The Dirac equation was integrated from $t=10$ to $t=-10$ using Mathematica's built-in function NDSolve. It took $\sim12-14$ minutes to compute $\Delta U$ on a laptop. After computing $\Delta U$ and $\Delta V$ for several values of $p_1=-p'_1$ and $p_2=p'_2=0$, and computing the inner products in~\eqref{N1compact}, we found a result that is to a good approximation just an overall constant times the spectrum in Fig.~\ref{fig:zGauss}. This is what one would expect from the instanton approach: For $p_2=p'_2=0$ we have an instanton with $y(u)=0$ that sees effectively a 2D field, $F_{\mu\nu}(x,y=0,t)$, which is identical to the field for Fig.~\ref{fig:zGauss}, which means that the exponential part of the probability is the same. This idea has also been discussed in~\cite{Linder:2015vta,DegliEsposti:2023qqu}.  

However, the pre-exponential factor is different. We have shown how to obtain the prefactor in several different cases~\cite{DegliEsposti:2022yqw,DegliEsposti:2023qqu,DegliEsposti:2024upq,DegliEsposti:2023fbv}. In each case, the first step is to derive a general formula for the prefactor that is suitable for numerical evaluation. We can use the same ideas to derive such a formula for general 3D (and 4D) fields with multiple peaks, but we have not yet done that\footnote{One could relatively quickly obtain a formula for the (exponentially) dominant contribution for symmetric 3D fields, as in~\cite{DegliEsposti:2023qqu} for 4D fields, and obtain a good approximation for the height, position and width of the spectrum. But for $E_0\sim1/3$ the subdominant contributions from lower field peaks ($E_\text{local max}<E_\text{global max}$) could give small but noticeable patterns on top of the dominant contribution. And for~\eqref{A0txy} there is a zero mode since $E(x,y)$ has a maximum on a circle, $x^2+y^2=\text{const.}\ne0$.}, so we leave a detailed comparison between the numerical and instanton results for 3D fields for future studies. Here we content ourselves with showing that it is possible to numerically solve the Dirac equation and study Schwinger pair production in 3D fields.

\section{Conclusions}

We have shown how to obtain the correlation between the momenta of the electron and positron produced in the Schwinger mechanism from solutions to the Dirac equation.  We have also shown how the Dirac equation can be solved for fields which depend on both time and space by separating the wave function into a background and a scattered wave. 

We have compared the results with the worldline-instanton approximation~\cite{DegliEsposti:2022yqw,DegliEsposti:2023qqu,DegliEsposti:2024upq} and found good agreement. In fact, in several examples the agreement is much better than what one would have expected, but that might be a coincidence. So far, the worldline approach has been used to obtain the leading order (LO) in a $E\ll1$ expansion, so one should expect relative errors on the order of $E$. But we do not want to consider too small $E$, because then the exponential suppression would make the results experimentally irrelevant. This leads us to consider $E\gtrsim0.1$ and we therefore expect relative errors of $\mathcal{O}(10\%)$. Of course, this argument only tells us how the error scales with $E$, not whether the coefficient is e.g. $2$ or $1/2$. The examples in this paper suggest that the error is not especially large even for $E=1/3$. By calculating the next-to-leading order (NLO) in the $E\ll1$ expansion, one would reduce the error to $\mathcal{O}(E^2)$.

However, we are not merely computing a single number. We are computing a function (the spectrum), and we have demonstrated that already the LO alone gives a good description of the shape of the spectrum. One can understand this by noting that the $E\ll1$ expansion of a term in the amplitude is expected to have the form
\be
M_j(E)\propto\sum a_{j,n} E^n e^{-b_j/E} \;,
\ee
where $a_n$ and $b$ depend on the other parameters but not on $E$, and where there will be several terms ($j=1,2,3...$) that give interference patterns for fields with multiple peaks. So far we have calculated $a_0$ and $b$. To obtain the NLO, we would also need $a_1$. But the shape of the spectrum is mostly determined by the real and imaginary parts of $b$. At least this is what we expect based on previous studies; see e.g. the comparisons between saddle-point and exact/numerical evaluations in Figs.~7-9 in~\cite{Dinu:2018efz} for nonlinear Compton scattering or Fig.~8 in~\cite{Dinu:2019pau} for nonlinear trident in plane-wave backgrounds, $A_1(t+z)$. But this is also confirmed by e.g. Figs.~\ref{fig:twoPeaksP} and~\ref{fig:twoPeaksDelta}. The height of each peak has a relative error as large as (or maybe even slightly smaller than) what one would expect from an error scaling as $E$. But the positions of the peaks and valleys have a much smaller error.  

Thus, the (LO) instanton approximation allows us to explore the parameter space much faster than by solving the Dirac equation numerically. This is particularly important for multidimensional fields, because the parameter space quickly becomes very large: In addition to the parameters describing the background field, the electron and positron momentum components are all independent for each nontrivial spatial dimension. It would therefore be natural to first scan the parameter space using the instanton approximation, and then, when one has found some interesting results, one can fix the precision by solving the Dirac equation numerically.     

One should of course keep in mind that both of these approaches only give the LO in the Furry picture. Radiation reaction, for example, is not included. RR is expected to be important for particles traveling at high energies on trajectories on which the electric and magnetic fields have non-negligible perpendicular components~\cite{DiPiazza:2011tq,Gonoskov:2021hwf,Fedotov:2022ely}, because those components become large after Lorentz boosting to the rest frame.     
However, for particles produced by the Schwinger mechanism, one could, in some cases, expect RR to be a small effect if the particles reach high energies by being accelerated parallel to the electric field and on trajectories with vanishing magnetic field~\cite{Brodin:2022dkd,DegliEsposti:2023qqu}. In any case, even if one could solve the Dirac equation to a high precision, one would probably not have a result that one could compare with some experiment to a high precision.    

Perhaps an even stronger motivation for solving the Dirac equation numerically would therefore be to check the instanton approximation, because it is not always obvious which instantons contribute. One usually obtains the correct instantons with a numerical continuation~\cite{Schneider:2018huk} starting at some simple field, for which one knows what to expect. For example, for~\eqref{SauterPulse} one could start at $\kappa=0$, where one can check the instanton result by comparing with other methods, and then gradually increase $\kappa$. When this works, each instanton will change shape gradually, and the resulting spectrum will also change gradually. Typically, we have one instanton for each peak of the field. The comparisons in this paper confirm the results obtained in this way. In~\cite{DegliEsposti:2024upq}, though, we found some additional instantons that might play a role in some parts of the momentum space further away from the peak of the spectrum. We leave an investigation of this for future studies.         

When planning for further studies, it is encouraging to find that the methods presented in this paper allow one to compute the spectrum quite quickly. At least for the simpler 2D fields ($E(t,z)$) considered here, each point in the spectrum only takes a couple of seconds or minutes to compute on the CPU on a laptop using Mathematica. We were also able to use this approach for 3D fields ($E(t,x,y)$), for which it typically took less than 20 minutes to solve the Dirac equation. So, if one first uses the instanton method to figure out what values of ${\bf p}$ and ${\bf p}'$ to consider, then even 3D fields are quite feasible, especially if one has access to a computer cluster and can compute several values of ${\bf p}$ and ${\bf p}'$ in parallel. 

However, one can expect this to be much faster on a GPU~\cite{BaukeGPU}. In fact, we have already confirmed this using the GPUs available online from Colab and the Diffrax solver~\cite{diffrax}. The code~\cite{GTGitHub} is still preliminary, but it can already reproduce the results that took $\mathcal{O}(10-20)$ minutes to compute on a laptop CPU in just a couple of seconds on a T4 GPU. Using the more powerful A100 GPU, it is even possible to handle 4D fields, $E(t,x,y,z)$, in a couple of minutes. We will further develop this approach and present details elsewhere.

\acknowledgements
Thanks to Fran\c{c}ois Fillion-Gourdeau, Holger Gies and Christian Kohlf\"urst for useful discussions, especially for explaining various features of other numerical approaches.
G. T. acknowledges support from the Swedish Research Council,
Contract No. 2020-04327.

\end{document}